# The many routes to AGN feedback

**Raffaella Morganti** [1,2,*]

[1]ASTRON, the Netherlands Institute for Radio Astronomy, Postbus 2, 7990 AA, Dwingeloo, The Netherlands.

[2]Kapteyn Astronomical Institute, University of Groningen, P.O. Box 800, 9700 AV Groningen, The Netherlands

**\* Correspondence:** morganti@astron.nl

## Abstract

The energy released by Active Galactic Nuclei (AGN) in the form of radiation, winds or radio plasma jets, is known to impact on the surrounding interstellar medium. The result of these processes, known as AGN (negative) feedback, is suggested to prevent gas, in and around galaxies, from cooling, and to remove, or at least redistribute, gas by driving massive and fast outflows, hence playing a key role in galaxy evolution. Given its importance, a large effort is devoted by the astronomical community to trace the effects of AGN on the surrounding gaseous medium and to quantify their impact for different types of AGN. This review briefly summarizes some of the recent observational results obtained in different wavebands, tracing different phases of the gas. I also summarise the new insights they have brought, and the constraints they provide to numerical simulations of galaxy formation and evolution. The recent addition of deep observations of cold gas and, in particular, of cold molecular gas, has brought some interesting surprises and has expanded our understanding of AGN and AGN feedback.

## 1    Introduction: AGN and feedback

AGN are fascinating objects. Since more than 50 years, astronomers have studied the effects of the extreme processes and the conditions that they cause. The huge amounts of energy released by the active super massive black hole (SMBH) can have an impact on the life and evolution of their entire host galaxy. This has made AGN relevant for an even broader community of astronomers. What is now commonly called *AGN feedback* has become a key ingredient in simulations of galaxy evolution. AGN feedback is the (self-regulating) process which links the energy released by the AGN to the surrounding gaseous medium and in this way, impacting on the evolution of the host galaxy. The energy injected by the AGN can provide the mechanism, either by preventing the cooling of gas or by expelling gas from the galaxy, to quench star formation (thus limiting the number of massive galaxies) and to limit the growth of the SMBH. Thus, it helps explaining key observations in these areas obtained in the recent years (e.g. Silk & Rees 1998; Gebhardt et al. 2000). In addition, the energy dumped in the environment of the AGN affects the fuelling of the nuclear activity itself, thereby regulating its duty-cycle.

AGN feedback is now included in many theoretical, numerical and semi-analytic models (e.g. Kauffmann & Haehnelt 2000; Di Matteo et al 2005; Schaye et al. 2015, Sijacki et al. 2015). The goal of the observations is to provide constraints to help a realistic implementation of feedback effects in these simulations. This is still a challenging task. As described below, the results from the increasingly accurate and deep observations have shown how complex these processes are.

Interestingly, the study of AGN in the context of their feedback effects has brought a number of unexpected discoveries (some of them described in this review) about the physical conditions of the gas in the surroundings of an active nucleus. Given its complexity, the treatment of feedback in numerical simulations is still extremely difficult. However, the implementation is becoming more sophisticated and reaching the stage where observables can be extracted from the simulations and compared with observations.

This review will concentrate on recent results from observations at low redshift and will focus on *negative feedback*. The energy released by the AGN and the related feedback effect can operate on very different distances from the central, active SMBH and can be traced by different phases of the gas. Particular emphasis will be given to the new information provided by the cold molecular gas. The review starts by summarising some of the results on the impact of the radio plasma on the hot and cold gas distributed on large (many tens to hundreds of kpc) scales. A more extended part of the review is dedicated to outflows occurring in the inner regions (kpc-scale). The mechanisms that may drive these outflows and their effects on the ISM are discussed. Finally, two examples of the diversity of mechanisms obtained thanks to multi-wavelengths observations will be presented.
It is impossible in this short paper to do justice to all exciting results. However, several other reviews have extensively covered various aspects and results of AGN feedback (see e.g. Cattaneo et al. 2009, Fabian 2012, Alexander & Hickox 2012, Harrison 2017) and the reader can refer to them for further information.
Before entering the description of the results from observations, it is important to note that, to first order, there are two modes in which AGN feedback can operate and they depend mostly on the type of nuclear activity (see Fabian 2012, Alexander & Hickox 2012, Harrison 2017 for reviews). Examples of manifestations of these modes are described in Sec. 2 and 3.

The *quasar (or radiative) mode* is mostly associated with high luminosity AGN, i.e. those emitting close to the Eddington limit, where most of the energy is released by radiation (or a wind) from the accretion disk but where also radio jets can play a role. The release of energy drives gas outflows expelling gas from the galaxy. The *jet (or kinetic) mode* is, instead, considered to be dominant in low-power AGN where the radio plasma provids the main source of energy, preventing the gaseous atmosphere from cooling back into the galaxy. While in the most extreme cases - high luminosity AGN and cool-core clusters – the separation between the two modes can be clear-cut, in other more intermediate situations, and for the most common types of AGN, the separation may not be always so sharp. Multi-wavelengths observations are often needed to recognise and disentangle different modes of feedback. Furthermore, the luminosity (or radio power) of the AGN is not the only parameter defining the impact. The coupling between the energy released and the medium is important (Tadhunter 2008, Alexander & Hickox 2012) as well as the duty-cycle of the activity (Fabian 2012, Morganti 2017).

## 2 Feedback on large (tens of kpc) scales: radio lobes and X-ray cavities

The most clear observational evidence of the impact of the nuclear activity on the surrounding medium comes from X-ray and radio observations, in particular in gas rich cool-core clusters (McNamara & Nulsen 2007, 2012 for reviews). X-ray images have revealed giant cavities and shock fronts in the hot gas, often filled with radio plasma lobes. The spatial coincidence between these cavities and the emission from radio lobes, suggests that the hot gas has been displaced by the expanding radio bubbles inflated by radio jets emitted by the central active SMBH. These studies have shown that the power associated to the radio jet provides the mechanism to offset radiative losses and to suppress gas cooling. They have provided a direct and relatively reliable mean of measuring the energy injected by the AGN (see McNamara & Nulsen 2007, 2012 for reviews). Given that the radiative cooling time at the centres of hot atmospheres in groups, clusters and elliptical galaxies can be short (ranging from < 1 Gyr to below 0.1 Gyr for the latter group), without this energy input, the gas would cool from the surrounding atmosphere. This will produce a large amount of molecular gas and star formation that instead is not observed in e.g. central cluster galaxies or any massive galaxy (Ciotti et al. 2017).

X-ray cavities are detected in about 50% of galaxy clusters, groups and individual galaxies. Their scales range from a few kpc to 200 kpc and the associated cavity power typically balances (or exceed) the X-ray cooling in galaxy clusters and also in some isolated elliptical galaxies (see Fig. 1 left). Furthermore, from the combined X-ray and radio studies, we have learned about the energetics involved**.** The mechanical power of radio jets largely exceeds their synchrotron power, suggesting that even weak radio sources - both in clusters and in elliptical galaxies – are mechanically powerful enough, Cavagnolo et al. (2010). Thus, this phenomenon appears to be important even in AGN of moderate radio luminosity. This is important because these AGN are much more common compared to the high power radio sources. Finally, it is worth to underline that the effect of the radio plasma can be relevant also in sources outside large clusters, as demonstrated by the work on isolated galaxies (e.g. Nulsen et al. 2007, Cavagnolo et al. 2010).

A small fraction (few %) of the most rapidly cooling gas in clusters is known to cool to low temperatures. This gas can provide the observed cold molecular gas reservoirs and feed the star formation in the central galaxy (Edge 2001, Salomé & Combes 2003). These initial findings are now, especially thanks to ALMA, expanded for more objects and they show a possible coupling between radio bubbles and cold gas. Fig. 1 (right) shows the case of the Phoenix cluster (Russell et al. 2017) where the extended filaments are draped around the expanding radio bubbles suggesting that the molecular gas flow is shaped by the radio-jet activity and possibly lifted by the radio bubbles. In the case of this cluster, the smooth velocity gradient observed in the filaments suggests an ordered flow with the gas velocities too low for the bulk of the cold gas to escape the galaxy. This gas will

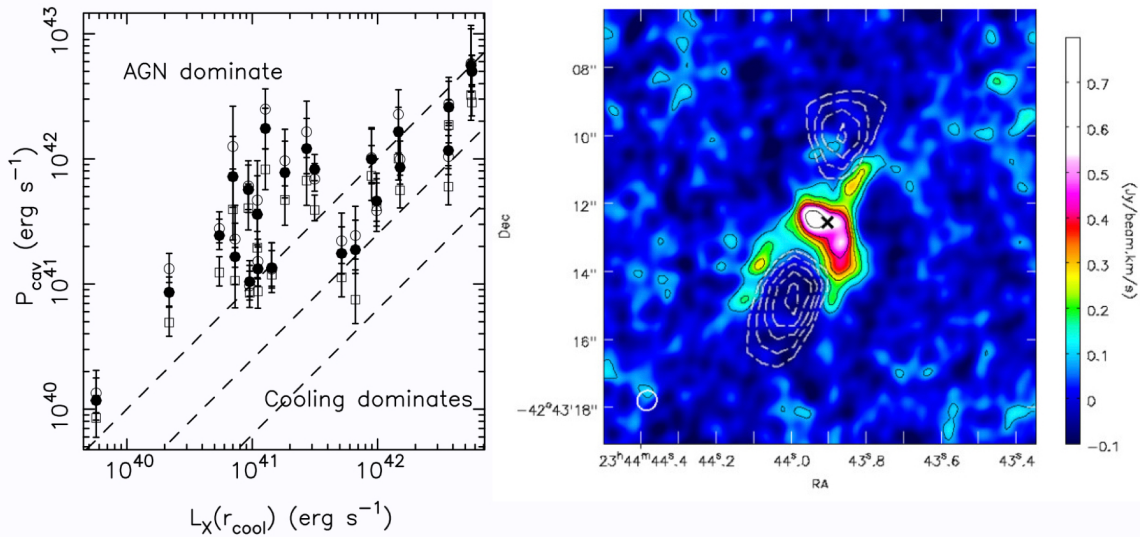

**Figure1** — *Left: Cavity heating power (estimated as $pV/t_a$, i.e. pressure, volume and age, see Nulsen 2009) plotted against the cooling power for objects ranging from luminous clusters, through groups, to nearby elliptical galaxies. The dashed lines represent the pV, 4pV and 16pV, from top to bottom; reproduced from Nulsen et al. (2009), with the permission of AIP Publishing). Right: CO(3-2) integrated intensity map for velocities 0 to +480 km s$^{-1}$ in the Phoenix Cluster. The X-ray cavities are shown by the dashed white contours. Image taken from Russell et al. (2017), © AAS. Reproduced with permission.*

eventually fall back toward the galaxy centre: this further support the (short) duty cycle of activity observed especially for cluster radio sources (Hogan et al. 2015). An increasing number of ALMA observations of molecular gas at the centres of clusters have now shown this synergy, with cold gas filaments extending along the radio bubbles (Russell et al. 2014, 2017; McNamara et al. 2014; Tremblay et al. 2016). This close coupling is essential to explain the self-regulation of feedback.

## 3     Effects of AGN on kpc scales: multi-phase gas outflows

Quenching star formation and stopping gas accretion onto the SMBH can also be achieved by massive outflows that can expel the gas from the central regions of galaxies. AGN-driven outflows, observed from pc to kpc scales, are known since a long time. The recent observations confirm that they are a relatively common phenomenon. Historically, they have been traced using mostly the ionised gas, observed in optical, UV and X-ray emission and absorption lines (see Veilleux et al. 2005, Bland-Hawthorn et al. 2007, Tadhunter 2008, King & Pounds 2015 for reviews). However, the most recent view that emerges from the continuously improving observations and by new or upgraded facilities - in particular radio and millimetre telescopes - is that the gas *outflows are truly complex and multi-phase*. The discovery of massive and fast outflows of cold gas (HI and cold molecular, Morganti et al. 2005, Feruglio et al. 2010) has taken everybody by surprise. Thus, different phases of the gas take part in the outflows and, in order to get the full picture of their physical properties, multi-wavelengths observations are needed. An overview of what found in term of mass outflow rate and kinetic comparison for the various components of the outflows, is presented in Tadhunter (2008). This shows how important is to trace and measure all these components. Furthermore, to fully gauge the impact of the outflows, both detailed single objects studies (see Sec. 5) as well as observations of large samples are required. Below we start with a summary of some of the main results obtained using the different gas tracers.

## 3.1 Ultra-Fast Outflows (UFO)

X-ray and UV observations have had, from the beginning, an important role in tracing outflows using absorption lines from ionized gas (King & Pounds 2015, Costantini 2010, Crenshaw et al. 2003, 2012a). About half of local Seyfert galaxies host a photoionized warm absorber (WA), which produces features detectable in soft X–ray (~0.3-2 keV) spectra and in the UV band. The latter absorption lines are usually blueshifted several hundred km s$^{-1}$ with respect of the systemic velocity, which indicates a global outflow of the absorbing gas with mass outflow rates in the range 0.01–0.06 M$_\odot$ yr$^{-1}$. These values are often comparable to the mass accretion rate but they provide only a small fraction of the bolometric luminosity, i.e. L$_{kin}$ < 0.1%L$_{bol}$ (Costantini 2010). Likely of higher impact are the highly ionized outflows with mildly relativistic velocities (i.e. $v \sim 0.1 - 0.25c$, where $c$ is the speed of light) traced by highly blue-shifted X–ray absorption lines in the iron K band (Pounds et al. 2003; Reeves et al. 2009, Tombesi et al. 2011, 2012). These ultrafast outflows are also known as UFO (Tombesi et al. 2011). A blind search carried out in archival spectra of the XMM-Newton archive has shown that they are present in about 35% of Seyfert galaxies. They differ from classical WA because of the higher outflow velocity and of the higher ionization. Indirect arguments indicate that their location is at sub-pc scales, i.e. ~ 0.0003 - 0.03 pc (see Reeves et al. 2009, Tombesi et al. 2011, 2012 and ref. therein).

Thus, these highly ionised flows are originated in the inner regions of the AGN outflows, and are likely driven by wide-angle winds launched from the accretion disks, accelerated by radiation pressure (e.g. Elvis 2000, King & Pounds 2003, Proga & Kallman 2004). Their mass outflow rate is in the range ~0.01 - 1 M$_\odot$ yr$^{-1}$ and the mechanical power is between ~$10^{42}$ and $10^{45}$ erg s$^{-1}$. These values typically correspond to >0.5% L$_{bol}$ (Tombesi et al. 2012, King & Pounds 2015). Important is the connection with the large scale outflows: the UFO likely represent the inner wind, which would then drive the massive molecular outflow (see below) on larger distances (see e.g. Tombesi et al. 2015, Feruglio et al. 2015, Veilleux et al. 2017), making them very relevant for feedback (see also Sec. 5).

## 3.2 Outflows of warm ionised gas on kpc scales

Outflows of warm (T ~ $10^4$ K) ionised gas have been observed in optical/IR lines and mostly traced, in low-redshifts objects, by the strong forbidden emission lines, e.g. [O III] $\lambda$5007 (see also discussion in Zakamska & Greene 2014). Outflows have been detected in ~30% of Seyfert galaxies (e.g. Crenshaw et al. 2003, 2007, 2012b), but they are common also in AGN of different luminosity selected from spectroscopic surveys (e.g. Zakamska & Greene 2014, Mullaney et al. 2013).

Radiative winds are considered often at the origin of these outflows (see below). However, other mechanisms are also observed. For example, radio plasma jets are known since early studies to affect the kinematics of the gas (e.g. Whittle 1985, Holt et al. 2008, Nesvadba et al. 2008 for some examples). Interestingly, a connection between the radio luminosity and the presence of outflows has been seen also in samples not selected based on radio properties and including radio weak (L$_{1.4 GHz}$= $10^{23}$-$10^{25}$ W Hz$^{-1}$, Mullaney et al. 2013) or even radio-quiet (<$10^{23}$

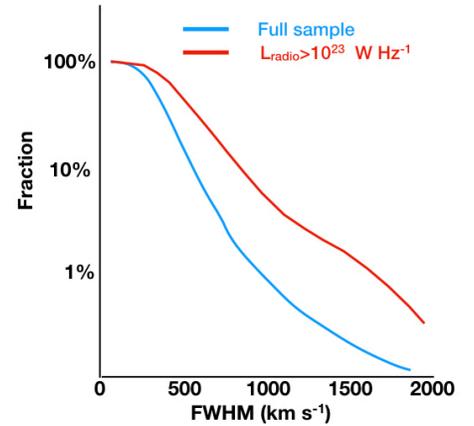

*Figure 2* Trend of the fraction FWHM of the [O III]$\lambda$5007 lines for the AGN in Mullaney et al.(2013). The cyan line represents the entire sample while the red line illustrates that objects with FWHM > 1000 km s$^{-1}$ are ≈5 times more prevalent among AGNs with L>$10^{23}$ W Hz$^{-1}$ compared to the overall AGN population. The figure is based on data from Mullaney et al (2013) and that no permission is required as the figure is original.

W Hz$^{-1}$, Zakamska & Greene 2014) objects. Figure 2 shows the trend found by Mullaney et al. (2013). In such weak radio sources, the radio emission is typically small or confined to faint radio jets. In the case of the faintest objects, the radio may even be a by-product of the shocks associated to the outflows, as suggested by Zakamska & Greene 2014.

The mass outflow rates associated with the outflows of warm ionised gas are usually found to be modest (at most a few M$_\odot$ yr$^{-1}$). However, these values are affected by uncertainties, in particular on the density of the outflowing gas (Holt et al. 2011) and on the spatial extent of the outflows (Husemann et al. 2013). The latter is still a matter of debate. The outflows of warm ionised gas are seen to extent up to ~ 10 kpc in the sample studied by Sun et al. (2017). On the other hand, for a sample of low-redshift QSOs, Husemann et al. (2013) found signatures of outflowing (>400 km s$^{-1}$) ionised gas on kpc scales only in 3 objects where a radio jet is most likely driving the outflow. These differences imply a wide range of energy efficiency for the outflows (e.g. $\eta = \dot{E}/L_{bol} = 0.01\% - 30\%$ estimated by Sun et al. 2017). It is also interesting to note the suggestion that outflows are driven by AGN episodes with ~ 10$^8$-year durations and shorter flickers 10$^6$ yr (Sun et al. 2017).

Finally, although outflows are common, it is worth notice that detailed work both with long slit and, more recently, with integral field units (IFU) shows that a complex interplay exists between outflows and infalling gas (e.g. Riffel et al. 2015), complicating in many objects the interpretation of the data.

### 3.3 The surprising presence of massive outflows of cold gas (HI and molecular)

The new "*entry*" in the study of AGN-driven outflows has been the discovery that a cold component of gas, i.e. in the neutral atomic (HI) and molecular (CO, OH) phases, can be associated even with fast outflows (with velocities ≥ 1000 km/s). This discovery has challenged our ideas on how the energy released by an AGN affects its surroundings. This component is particularly interesting for two main reasons: its uncertain origin (see Sec 4) - perhaps the result of a very efficient cooling of the gas - and its impact. Surprisingly, this component (and in particular the cold molecular component) has been found to represent the most massive component of the outflow.

An extensive literature is available on the properties of these outflows: for the atomic gas by e.g., Rupke & Veilleux 2011, 2013; Lehnert et al. 2011; Morganti et al. 2005, 2013a, 2016; for the warm and cold molecular gas by e.g., Feruglio et al. 2010; Alatalo et al. 2011, Dasyra & Combes 2011, 2012; Guillard et al. 2012; García-Burillo et al. 2014; Tadhunter et al. 2014; Cicone et al. 2014; Morganti et al. 2015a; and for OH by e.g., Fischer et al. 2010; Sturm et al. 2011; Veilleux et al. 2013.

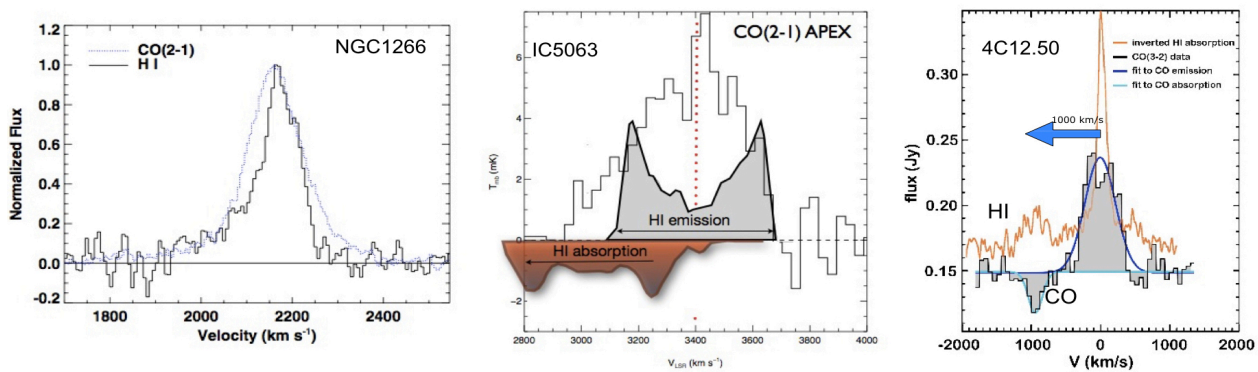

*Figure 3* – Three examples of HI and CO integrated line profiles suggesting similar kinematical properties of the two phases of the gas. From left to right: NGC1266 (figure taken from Alatalo et al. 2011, © AAS. Reproduced with permission), IC5063 (figure taken from Morganti et al. 2013b, reproduced with permission © ESO) and 4C12.50 (Figure taken from Dasyra & Combes 2010, reproduced with permission © ESO).

Figure 3 shows the intriguing similarities between the CO and HI profiles for three objects suggesting similar kinematics for these components. It is not yet clear how far this synergy extends, given the limited number of objects with HI and CO observations available.

Using data from literature, Cicone et al. (2014) and Fiore et al. (2017) have shown correlations between the AGN-driven molecular and ionized-wind mass outflow rates and the AGN bolometric luminosity. The mass outflow rates of the molecular gas are systematically the highest, except for the most luminous AGN where the two rates appear to converge (Fiore et al. 2017). A correlation is also found between the momentum carried in the outflow ($v\mathrm{d}M/\mathrm{d}t$, where $v$ is the outflow velocity) with the photon momentum output of the AGN, $L_{AGN}/c$ but with a "momentum boosting" of about a factor 20, suggesting an energy-driven nature of the outflows (see e.g. Cicone et al. 2014).

Fast HI outflows are observed using HI absorption. The planned upcoming surveys (Morganti et al. 2015b) make this method quite powerful for investigating the presence of these outflows in large samples. A proof-of-concept survey is presented in Gereb et al. (2015) and Maccagni et al. (2017a). Of the galaxies detected in HI, at least 15% show fast outflows (Gereb et al. 2015) with mass outflow rates between a few and ~ 30 $M_\odot$ yr$^{-1}$. The radio jets drive at least some of these outflows. In particular, young radio sources are those where the plasma jet has the strongest impact on the surrounding medium and where most of the HI outflows are occurring (Gereb et al. 2015). A possible correlation has been observed between the amplitude (in velocity) of the outflow and the radio power (Maccagni et al. 2017a). Finally, the presence of clumpy gas medium in the regions surrounding the radio sources may enhance the impact of the jet (as suggested by some simulations, see Sec. 4). The presence of such a clumpy medium (embedded in a diffuse component) is confirmed by the results of deep, high-resolution (VLBI) observations tracing HI clouds with tens of pc sizes (with cloud masses up to ~ $10^4$ $M_\odot$, Morganti et al. 2013, Schulz et al. in prep). Molecular clouds are also traced by high resolution ALMA observations (Maccagni et al. 2017 in prep).

## 4    Driving mechanisms and energetics of the outflows

Different mechanisms have been proposed to drive the outflowing gas (see e.g. Veilleux et al. 2005, King & Pounds 2015). Wide-angle winds can be launched from the accretion disk and driven by the coupling of the radiation to the ambient medium through radiation pressure on dust. It is also possible to have a hot thermal wind (e.g., Compton-heated) colliding with and accelerating the ISM. Parsec-scale jets can produce over-pressured cavities from which the wide-angled outflow can be launched. Alternatively, outflows can be driven by the mechanical action of the radio plasma emanating from the AGN (Wagner et al. 2012, Mukherjee et al. 2016).

For powerful AGN, the scenario that has been suggested to better match the observations - in particular the presence of cold, molecular gas - is the one where winds launched from the accretion disk interact and shock the surrounding medium. This interaction can create an (energy-conserving) adiabatically expanding hot bubble. The adiabatically expanding wind sweeps up gas and drives an outer shock into the host ISM (Zubovas & King, 2012, 2014; Faucher-Giguère & Quataert 2012). The outflowing gas is able to cool radiatively into clumps of cold molecular material. This would explain the presence of fast outflows (~1300 km s$^{-1}$) of neutral-atomic (HI, NaI D) and molecular gas (Costa et al. 2017, Richings & Faucher-Giguère 2017). An alternative scenario to explain the fast outflows of molecular gas assumes that pre-existing molecular clouds from the host ISM are entrained in the adiabatically expanding shocked wind and they can be accelerated to the observed velocities without being destroyed (see e.g. Scannapieco 2017).

The role of radio jets is also relevant. In this case, the structure of the medium is playing an important role in defining the impact that the radio plasma jets can have (e.g. Wagner et al. 2012). Following the results of recent simulations, a jet entering a clumpy medium can have a larger impact than previously proposed. Numerical simulations of a newly created radio jet entering a two-phase clumpy medium (dense molecular clouds, i.e. a few hundred cm$^3$, embedded in a lower-density medium) show that the jet expands following a path of less resistance but still colliding with the gas clouds. This originates a cocoon of disturbed and outflowing gas around the jet, affecting a *large* region of the galaxy (see simulations by Wagner et al. 2012 and Mukherjee et al. 2016).

The requirements from the simulations in order to explain the observations are such that the kinetic power in the wind has to be a substantial fraction of the available accretion power ($\dot{E}/L_{edd}$ ~0.05 - 0.1, Di Matteo et al. 2005). Although uncertainties still affect some key parameters of the outflows (like e.g. mass, mass outflow rate etc.), the results so far suggest that these requirements are not always met. Alternatively, the coupling between the energy released and the surrounding medium has to be particularly efficient, i.e. as a consequence of the conditions of the medium. For example, and as mentioned above for the case of radio jets, an important factor for the coupling of photons or plasma with the multiphase galactic gas is the presence of a clumpy medium with dense clouds (Wagner et al. 2012, Bieri et al. 2017). This is important both for radiation-driven as for the jet-driven outflows (Wagner et al. 2012, 2013). Furthermore, a "two stage" model has been proposed by Hopkins & Elvis (2010) in which the outflow passing over a cold cloud, will be able to affect the cloud material and expand it in the direction perpendicular to the outflow direction. This shredding/expansion will amplify the effect of the interaction and substantially reduce (i.e. up to an order of magnitude) the required energy budget for feedback to affect a host galaxy.
Finally, connecting outflows observed on different scales, i.e. connecting the inner wind with large-scale molecular outflows, is important for obtaining a global view of outflows. This has been possible so far only for a limited number of objects. These cases suggest that most of the nuclear wind kinetic energy produced on the pc scale is transferred to mechanical energy on the kpc scale outflow. At least in some cases, the outflow is undergoing an energy conserving expansion (Cicone et al. 2014, Tombesi et al. 2015, Stern et al. 2016) although more detailed data are needed to confirm this scenario (see e.g. Veilleux et al. 2017).

## 5 The variety of outflows through two case studies

Despite these exciting results, the number of objects where extensive multi-wavelengths data are available to characterize the different phases of the outflow is still limited. This is mainly because of the time and effort needed to collect data from a variety of (often highly oversubscribed) telescopes. Expanding this limited statistic and connecting the different scales and phases of the outflowing gas is the goal of many on-going projects. Below, I summarise the results for two of the best-studied objects. They represent cases where, thanks to detailed observations, two different mechanisms for driving the outflows have been suggested.

### 5.1 The wind-driven outflow in Mrk 231

Mrk 231 represents one of the best examples of wind-driven outflow and the first case where fast outflowing molecular gas was observed (Fisher et al. 2010, Feruglio et al. 2010). Being the nearest quasar, it allows a detailed exploration of the physical conditions of the gas in the outflow. A nuclear UFO has been observed in X-ray with velocity 20000 km s$^{-1}$ and mass outflow rate in the range 0.3 - 1.6 $M_\odot$ yr$^{-1}$ (Feruglio et al. 2015). On the sub-kpc scale, an HI outflow of ~1300 km s$^{-1}$ has been detected (Morganti et al. 2016) while the molecular outflow extend to kpc scale with a mass outflow rate in the range 500 -1000 $M_\odot$ yr$^{-1}$ and associated kinetic energy of $E_{kin,mol}$ = [7 -10] x 10$^{43}$ erg s$^{-1}$ (Feruglio et al. 2015). The NaI outflow is the most extended, reaching 3 kpc and mass outflow rate of 179 $M_\odot$ yr$^{-1}$ (Rupke et al. 2011, 2013). This complex outflow is explained as driven by a wide angle, wind (see Fig.17 in Feruglio et al. 2015). Although Mrk 231 contains a radio plasma jet and radio lobes, the deep and detailed observations available show that the jet power does not seem large enough to drive and sustain the outflow. Mrk 231 is one of the few objects where the parameters of the outflow on different scales have been connected. Interestingly $E_{kin,UFO}$ ~ $E_{kin,mol}$ as expected for outflows undergoing an energy conserving expansion. Thus, this confirms that most of the UFO kinetic energy is transferred to mechanical energy of the kpc-scale outflow (Feruglio et al. 2015).

## 5.2 The jet-driven outflow in the Seyfert 2 IC 5063

IC 5063 represents one of the most radio-loud Seyfert galaxies (albeit still a relatively weak radio AGN in a general sense; $P_{1.4\ GHz}$ = 3×10$^{23}$ W Hz$^{-1}$) and it was the first object where a fast AGN-driven HI outflow was discovered (Morganti et al. 1998, Oosterloo et al. 2000). The outflow is multi-phase (including HI, ionized, warm and cold molecular gas, see Morganti et al. 2015 for a summary). Although the radio power is actually lower than the one of Mrk 231, the location of the outflow suggests that in IC5063 the radio jet is the dominant driving mechanism of the outflow (Tadhunter et al. 2014, Morganti et al. 2015a). ALMA observations of the molecular gas show that disturbed kinematics of the gas occur across the entire extension of the radio source (~1 kpc, see Fig. 4 left). Furthermore, not only the kinematics is affected by the jet but also the excitation of the gas, with the outflowing gas having high excitation (with $T_{ex}$ 30-50 K) and optically thin conditions compared to the gas in the quiescent disk (see Fig. 4 right; Dasyra et al. 2016, Oosterloo et al. 2017). The mass of the cold molecular outflow ~ 1.2 × 10$^6$ $M_\odot$ is much higher than the one associated to warm H$_2$ and to ionised gas. This suggests that most of the cold molecular outflow is due to fast cooling after the passage of a shock. The mass outflow rate of the cold molecular gas is ~ 4 $M_\odot$yr$^{-1}$ (while for the ionised gas is ~ 0.08 $M_\odot$yr$^{-1}$). In this object, the kinetic power of the outflow appears to be a relative high fraction of the nuclear bolometric luminosity. However, the global impact is modest and only a small fraction of the outflowing gas may leave the galaxy. The main effect of the outflow is to increase the turbulence of the medium and redistribute the gas (see e.g. Guillard et al. 2012, 2015).

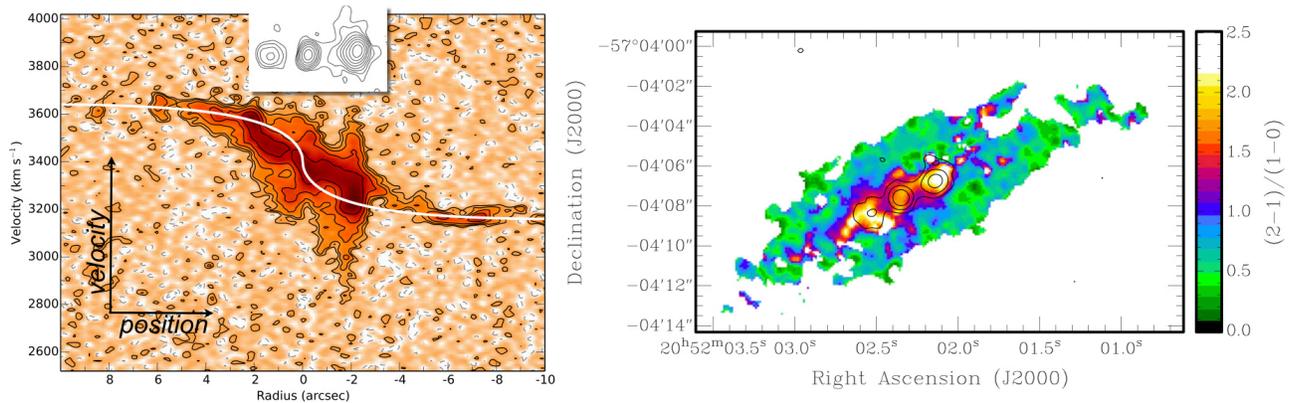

***Figure 4*** *– ALMA observations of IC 5063. (Left) integrated position-velocity map of the CO(2-1) along the radio axis of IC 5063. The white line indicates the expected kinematics of gas following regular rotation (figure modified from Morganti et al. 2015a, reproduced with permission © ESO). (Right) Ratio CO(2-1)/CO(1-0) integrated brightness temperatures I (in K kms$^{-1}$), courtesy of T. Oosterloo. Overplotted are the contours of the 346 GHz continuum emission of IC 5063 (see also Oosterloo et al. 2017).*

## 6      Summary and conclusions

As described in this review, there are a variety of ways in which the energy released by the AGN affects the surrounding medium.

On large scales, radio plasma lobes excavate cavities in the hot gas depositing enough energy to prevent this large reservoir of gas from cooling. The X-ray cavities are detected in ~50% of galaxy clusters, groups and individual galaxies. Some of the gas manages to cool and can be followed to trace its interaction with the radio plasma. New exciting possibilities to explore this phenomenon are open up by ALMA. The velocities of the cold molecular gas observed so far are not high enough for the gas to escape. The gas would fall back, further supporting the (short) duty cycle of activity observed especially for radio sources in cool-core clusters (Hogan et al. 2015).

Closer to the nucleus, outflows are found in a large fraction of AGN while, at the same time, infall is also required to maintain the AGN active (Kurosawa et al. 2009): a complex interplay – still to be fully disentangled - is occurring in these central regions. The gas participating to the outflows is multi-phase and includes also a cold (HI and cold molecular) component. Thus, different tracers can be used to study the physical properties of the outflows and the driving mechanisms. Surprisingly, the component of cold molecular gas is often the most massive and, therefore, represents an important component for feedback. The mechanism(s) that can produce fast outflows of cold (molecular) gas are still uncertain. This gas could be the result of fast cooling after the passage of a shock: the cold molecular gas would be the end product of the cooling. However, alternative scenarios have been suggested and need to be verified with more data. The kinetic energy of the outflows is not always matching the requirements from numerical simulations but more sophisticated models of the coupling between the energy and the medium is needed to fully assess the impact. The effect of radio jets can be important, also for low power radio sources. Interestingly, the initial phase of a radio jet (i.e. in young or restarted radio source) appears to have the largest impact on the surrounding gas. Also in the case of the outflows, the gas does not always escape the galaxy:    the main effect of the AGN appears to be to inject turbulence and relocate the gas.

This review has focused on AGN at low redshift. However, the results presented here can provide an important reference point for the studies at high-z where the effects of AGN feedback are expected to be larger. Interesting results for high redshift objects are now starting to appear (e.g. Carniani et al. 2017). Finally, this review has focused on *negative* feedback but *positive* feedback is also expected to occur (see e.g. Maiolino et al. 2017). At low redshift this effect, albeit observed, seems to be limited (Santoro et al. 2016 and ref. therein, see also Combes 2017, these Proceedings) but more dedicated observations are needed to quantify its importance.

In summary, the field is expanding extremely rapidly and, thanks to the many upcoming new observing facilities, our view of AGN will become increasingly detailed. An even tighter collaboration between observers and theorists will be necessary to interpret this wealth of data for understanding the complex phenomenon of AGN feedback.

*The authors declare that the research was conducted in the absence of any commercial or financial relationships that could be construed as a potential conflict of interest.*

## 7      Acknowledgments

RM would like to thank the organizers for the interesting and pleasant conference and for giving the chance to the participants to visit unique places in Padova, taking advantage of the 250th anniversary of the Astronomical Observatory (la Specola). RM gratefully acknowledges support from the European Research Council under the European Union's Seventh Framework Programme (FP/2007-2013) /ERC Advanced Grant RADIOLIFE-320745.

# 8   Reference

Alatalo, K., Blitz, L., Young, L. M., Davis, T. A., Bureau, M., Lopez, L. A., et al. 2011. Discovery of an Active Galactic Nucleus Driven Molecular Outflow in the Local Early-type Galaxy NGC 1266. The Astrophysical Journal 735, 88-100.

Alexander, D. M., Hickox, R. C. 2012. What drives the growth of black holes?. New Astronomy Reviews 56, 93-121.

Bland-Hawthorn, J., Veilleux, S., & Cecil, G. 2007, Galactic winds: a short review. Astrophysics and Space Science 311, 87-98.

Bieri, R., Dubois, Y., Rosdahl, J., Wagner, A., Silk, J., Mamon, G.A. 2017. Outflows driven by quasars in high-redshift galaxies with radiation hydrodynamics. Monthly Notices of the Royal Astronomical Society 464, 1854-1873.

Carniani, S., Marconi, A., Maiolino, R., Feruglio, C., Brusa, M., Cresci, G. et al., 2017, AGN feedback on molecular gas reservoirs in quasars at z 2.4. Astronomy and Astrophysics 605, 105-119.

Cattaneo, A., Faber, S. M., Binney, J., Dekel, A., Kormendy, J., Mushotzky, R., et al. 2009. The role of black holes in galaxy formation and evolution. Nature 460, 213-219.

Cavagnolo, K.W., McNamara, B.R., Nulsen, P.E.J., Carilli, C.L., Jones, C., Bîrzan, L. 2010. A Relationship Between AGN Jet Power and Radio Power. The Astrophysical Journal 720, 1066-1072.

Cicone, C., Maiolino, R., Sturm, E., Graciá-Carpio, J., Feruglio, C., Neri, R.; et al. 2014, Massive molecular outflows and evidence for AGN feedback from CO observations. Astronomy and Astrophysics 562, 21-46.

Ciotti, L., Pellegrini, S., Negri, A., Ostriker, J.P. 2017. The Effect of the AGN Feedback on the Interstellar Medium of Early-Type Galaxies:2D Hydrodynamical Simulations of the Low-Rotation Case. The Astrophysical Journal 835, 15-37.

Combes, F. 2017. AGN feedback and its quenching efficiency. Frontiers in Astronomy and Space Sciences 4, 10.

Costa, T., Rosdahl, J., Sijacki, D., Haehnelt, M.G. 2017. Quenching star formation with quasar outflows launched by trapped IR radiation. ArXiv e-prints arXiv:1709.08638.

Costantini, E. 2010. The Ultraviolet-X-Ray Connection in AGN Outflows. Space Science Reviews 157, 265-277.

Crenshaw, D.M., Fischer, T.C., Kraemer, S.B., Schmitt, H.R., Turner, T.J. 2012b. Measuring Feedback in Nearby AGN. AGN Winds in Charleston, ASPC 460, 261-265.

Crenshaw, D.M., Kraemer, S.B. 2012a. Feedback from Mass Outflows in Nearby Active Galactic Nuclei. I. Ultraviolet and X-Ray Absorbers. The Astrophysical Journal 753, 75-86.

Crenshaw, D.M., Kraemer, S.B. 2007 Mass Outflows from Seyfert Galaxies as Seen in Emission and Absorption. The Central Engine of Active Galactic Nuclei 373, 319-328.

Crenshaw, D.M., Kraemer, S.B., George, I.M. 2003. Mass Loss from the Nuclei of Active Galaxies. Annual Review of Astronomy and Astrophysics 41, 117-167.

Dasyra, K.M., Combes, F., Oosterloo, T., Oonk, J.B.R., Morganti, R., Salomé, P., Vlahakis, N. 2016. ALMA reveals optically thin, highly excited CO gas in the jet-driven winds of the galaxy IC 5063. Astronomy and Astrophysics 595, L7-L11.

Dasyra, K.M., Combes, F. 2012. Cold and warm molecular gas in the outflow of 4C 12.50. Astronomy and Astrophysics 541, L7-L11.

Dasyra, K.M., Combes, F. 2011. Turbulent and fast motions of $H_2$ gas in active galactic nuclei. Astronomy and Astrophysics 533, L10-L14.

Di Matteo, T., Springel, V., Hernquist, L. 2005. Energy input from quasars regulates the growth and activity of black holes and their host galaxies. Nature 433, 604-607.

Edge, A.C. 2001. The detection of molecular gas in the central galaxies of cooling flow clusters. Monthly Notices of the Royal Astronomical Society 328, 762-782.

Elvis, M. 2000. A Structure for Quasars. The Astrophysical Journal 545, 63-76.

Fabian, A.C. 2012. Observational Evidence of Active Galactic Nuclei Feedback. Annual Review of Astronomy and Astrophysics 50, 455-489.

Faucher-Giguère, C.-A., Quataert, E. 2012. The physics of galactic winds driven by active galactic nuclei. Monthly Notices of the Royal Astronomical Society 425, 605-622.

Feruglio, C., Maiolino, R., Piconcelli, E., Menci, N., Aussel, H., Lamastra, A., Fiore, F. 2010. Quasar feedback revealed by giant molecular outflows. Astronomy and Astrophysics 518, L155-159.

Feruglio, C.; Fiore, F.; Carniani, S.; Piconcelli, E.; Zappacosta, L.; Bongiorno, A. et al. 2015. The multi-phase winds of Markarian 231: from the hot, nuclear, ultra-fast wind to the galaxy-scale, molecular outflow. Astronomy and Astrophysics 583, 99-115.

Fiore, F.; Feruglio, C.; Shankar, F.; Bischetti, M.; Bongiorno, A.; Brusa, M. et al. 2017.\ AGN wind scaling relations and the co-evolution of black holes and galaxies.\ Astronomy and Astrophysics 601, 143-164.

Fischer, J.; Sturm, E.; González-Alfonso, E.; Graciá-Carpio, J.; Hailey-Dunsheath, S.; Poglitsch, A.; 2010. Herschel-PACS spectroscopic diagnostics of local ULIRGs: Conditions and kinematics in Markarian 231. Astronomy and Astrophysics 518, L41-45.

García-Burillo, S., Combes, F.; Usero, A.; Aalto, S.; Krips, M.; Viti, S et al. 2014. Molecular line emission in NGC 1068 imaged with ALMA. I. An AGN-driven outflow in the dense molecular gas.\ Astronomy and Astrophysics 567, 125-149.

**The many routes to AGN feedback**Gebhardt, K., Bender, Ralf; Bower, Gary; Dressler, Alan; Faber, S. M.; Filippenko, Alexei V. et al. 2000. A Relationship between Nuclear Black Hole Mass and Galaxy Velocity Dispersion. The Astrophysical Journal 539, L13-L16.

Geréb, K., Maccagni, F.M., Morganti, R., Oosterloo, T.A. 2015. The HI absorption ``Zoo". Astronomy and Astrophysics 575, 44-61.

Guillard, P., Boulanger, F., Lehnert, M.D., Pineau des Forêts, G., Combes, F., Falgarone, E., Bernard-Salas, J. 2015. Exceptional AGN-driven turbulence inhibits star formation in the 3C 326N radio galaxy. Astronomy and Astrophysics 574, 32-47.

Guillard, P., Ogle, P.M., Emonts, B.H.C., Appleton, P.N., Morganti, R., Tadhunter, C., Oosterloo, T., Evans, D.A., Evans, A.S. 2012. Strong Molecular Hydrogen Emission and Kinematics of the Multiphase Gas in Radio Galaxies with Fast Jet-driven Outflows. The Astrophysical Journal 747, 95-120.

Harrison, C.M. 2017. Impact of supermassive black hole growth on star formation. Nature Astronomy 1, 0165.

Harrison, C.M., Alexander, D.M., Mullaney, J.R., Swinbank, A.M. 2014. Kiloparsec-scale outflows are prevalent among luminous AGN: outflows and feedback in the context of the overall AGN population. Monthly Notices of the Royal Astronomical Society 441, 3306-3347.

Hogan, M. T., Edge, A. C., Geach, J. E., Grainge, K. J. B., Hlavacek-Larrondo, J., Hovatta, T., et al. 2015. High radio-frequency properties and variability of brightest cluster galaxies. Monthly Notices of the Royal Astronomical Society 453, 1223-1240.

Holt, J., Tadhunter, C.N., Morganti, R., Emonts, B.H.C. 2011 The impact of the warm outflow in the young (GPS) radio source and ULIRG PKS 1345+12 (4C 12.50). Monthly Notices of the Royal Astronomical Society 410, 1527-1536.

Holt, J., Tadhunter, C.N., Morganti, R. 2008. Fast outflows in compact radio sources: evidence for AGN-induced feedback in the early stages of radio source evolution. Monthly Notices of the Royal Astronomical Society 387, 639-659.

Hopkins, P.F., Elvis, M. 2010. Quasar feedback: more bang for your buck.\ Monthly Notices of the Royal Astronomical Society 401, 7-14.

Husemann, B., Wisotzki, L., Sánchez, S.F., Jahnke, K. 2013. The properties of the extended warm ionised gas around low-redshift QSOs and the lack of extended high-velocity outflows. Astronomy and Astrophysics 549, 43-76.

Kauffmann, G., Haehnelt, M. 2000. A unified model for the evolution of galaxies and quasars. Monthly Notices of the Royal Astronomical Society 311, 576-588.

King, A., Pounds, K. 2015. Powerful Outflows and Feedback from Active Galactic Nuclei.\ Annual Review of Astronomy and Astrophysics 53, 115-154.

King, A.R., Pounds, K.A. 2003. Black hole winds. Monthly Notices of the Royal Astronomical Society 345, 657-659.
13

Kurosawa R., Proga D., Nagamine K., 2009, ApJ, 707, 823

Lehnert, M.D., Tasse, C., Nesvadba, N.P.H., Best, P.N., van Driel, W. 2011. The Na D profiles of nearby low-power radio sources: jets powering outflows.\ Astronomy and Astrophysics 532, L3-L8.

Maiolino, R., Maiolino, R.; Russell, H. R.; Fabian, A. C.; Carniani, S.; Gallagher, R.; Cazzoli, S.; et al. 2017. Star formation inside a galactic outflow. Nature 544, 202-206.

Maccagni, F.M., Morganti, R., Oosterloo, T.A., Geréb, K., Maddox, N. 2017a. Kinematics and physical conditions of HI in nearby radio sources. The last survey of the old Westerbork Synthesis Radio Telescope. Astronomy and Astrophysics 604, 43-65.

McNamara, B.R., Russell, H. R., Nulsen, P. E. J., Edge, A. C., Murray, N. W., Main, R. A. et al. 2014. A $10^{10}$ Solar Mass Flow of Molecular Gas in the A1835 Brightest Cluster Galaxy. The Astrophysical Journal 785, 44.

McNamara, B.R., Nulsen, P.E.J. 2012. Mechanical feedback from active galactic nuclei in galaxies, groups and clusters. New Journal of Physics 14, 055023.

McNamara, B.R., Nulsen, P.E.J. 2007. Heating Hot Atmospheres with Active Galactic Nuclei. Annual Review of Astronomy and Astrophysics 45, 117-175.

Morganti, R. 2017. Archaeology of active galaxies across the electromagnetic spectrum.\ Nature Astronomy 1, 596-605.

Morganti, R., Veilleux, S., Oosterloo, T., Teng, S.H., Rupke, D. 2016. Another piece of the puzzle: The fast H I outflow in Mrk 231. Astronomy and Astrophysics 593, 30-41.

Morganti, R., Oosterloo, T., Oonk, J.B.R., Frieswijk, W., Tadhunter, C. 2015a. The fast molecular outflow in the Seyfert galaxy IC 5063 as seen by ALMA. Astronomy and Astrophysics 580, 1-12

Morganti, R., Sadler, E.M., Curran, S. 2015b. Cool Outflows and HI absorbers with SKA. Advancing Astrophysics with the Square Kilometre Array (AASKA14) 134. Online at http://pos.sissa.it/cgi-bin/reader/conf.cgi?confid=215, id.134

Morganti, R., Fogasy, J., Paragi, Z., Oosterloo, T., Orienti, M. 2013a. Radio Jets Clearing the Way Through a Galaxy: Watching Feedback in Action. Science 341, 1082-1085.

Morganti, R., Frieswijk, W., Oonk, R.J.B., Oosterloo, T., Tadhunter, C. 2013b. Tracing the extreme interplay between radio jets and the ISM in IC 5063. Astronomy and Astrophysics 552, L4-L9.

Morganti, R., Tadhunter, C.N., Oosterloo, T.A. 2005. Fast neutral outflows in powerful radio galaxies: a major source of feedback in massive galaxies. Astronomy and Astrophysics 444, L9-L13.

Morganti, R., Oosterloo, T., Tsvetanov, Z. 1998. A Radio Study of the Seyfert Galaxy IC 5063: Evidence for Fast Gas Outflow. The Astronomical Journal 115, 915-927.

Mukherjee, D., Bicknell, G.V., Sutherland, R., Wagner, A. 2016. Relativistic jet feedback in high-redshift galaxies - I. Dynamics. Monthly Notices of the Royal Astronomical Society 461, 967-983.

Mullaney, J.R., Alexander, D.M., Fine, S., Goulding, A.D., Harrison, C.M., Hickox, R.C. 2013. Narrow-line region gas kinematics of 24 264 optically selected AGN: the radio connection. Monthly Notices of the Royal Astronomical Society 433, 622-638.

Nesvadba, N.P.H., Lehnert, M.~D., De Breuck, C., Gilbert, A.M., van Breugel, W. 2008. Evidence for powerful AGN winds at high redshift: dynamics of galactic outflows in radio galaxies during the ``Quasar Era''. Astronomy and Astrophysics 491, 407-424.

Nulsen, P., Jones, C., Forman, W., Churazov, E., McNamara, B., David, L., Murray, S. 2009. Radio Mode Outbursts in Giant Elliptical Galaxies. American Institute of Physics Conference Series 1201, 198-201.

Oosterloo, T., Oonk, J.B.R., Morganti, R., Combes, F., Dasyra, K., Salomé, P., Vlahakis, N., Tadhunter, C. 2017. Properties of the molecular gas in the fast outflow in the Seyfert galaxy IC 5063.

Oosterloo, T.A., Morganti, R., Tzioumis, A., Reynolds, J., King, E., McCulloch, P., Tsvetanov, Z. 2000. A Strong Jet-Cloud Interaction in the Seyfert Galaxy IC 5063: VLBI Observations. The Astronomical Journal 119, 2085-2091.

Pounds, K.A., Reeves, J.N., King, A.R., Page, K.L., O'Brien, P.T., Turner, M.J.L. 2003. A high-velocity ionized outflow and XUV photosphere in the narrow emission line quasar PG1211+143. Monthly Notices of the Royal Astronomical Society 345, 705-713.

Proga, D., Kallman, T.R. 2004. Dynamics of Line-driven Disk Winds in Active Galactic Nuclei. II. Effects of Disk Radiation. The Astrophysical Journal 616, 688-695.

Reeves, J.N., O'Brien, P.T., Braito, V., Behar, E., Miller, L., Turner, T.J., Fabian, A.C., Kaspi, S., Mushotzky, R., Ward, M. 2009. A Compton-thick Wind in the High-luminosity Quasar, PDS 456. The Astrophysical Journal 701, 493-507.

Richings, A.J., Faucher-Giguere, C.-A. 2017. Radiative cooling of swept up gas in AGN-driven galactic winds and its implications for molecular outflows. ArXiv e-prints arXiv:1710.09433.

Riffel, R.A., Storchi-Bergmann, T., Riffel, R. 2015. Feeding versus feedback in active galactic nuclei from near-infrared integral field spectroscopy - X. NGC 5929. Monthly Notices of the Royal Astronomical Society 451, 3587-3605.

Rupke, D.S.N., Veilleux, S. 2013. The Multiphase Structure and Power Sources of Galactic Winds in Major Mergers. The Astrophysical Journal 768, 75.

Rupke, D.S.N., Veilleux, S. 2011. Integral Field Spectroscopy of Massive, Kiloparsec-scale Outflows in the Infrared-luminous QSO Mrk 231. The Astrophysical Journal 729, L27.

Russell, H.R., McDonald, M.; McNamara, B. R.; Fabian, A. C.; Nulsen, P. E. J.; Bayliss, M. B.; et al. 2017. Alma Observations of Massive Molecular Gas Filaments Encasing Radio Bubbles in the Phoenix Cluster. The Astrophysical Journal 836, 130

Russell, H.R., McNamara, B. R.; Edge, A. C.; Nulsen, P. E. J.; Main, R. A.; Vantyghem, A. N. et al. 2014. Massive Molecular Gas Flows in the A1664 Brightest Cluster Galaxy. The Astrophysical Journal 784, 78.

Salomé, P., Combes, F. 2003. Cold molecular gas in cooling flow clusters of galaxies. Astronomy and Astrophysics 412, 657-667.

Santoro, F., Oonk, J.B.R., Morganti, R., Oosterloo, T.A., Tadhunter, C. 2016. Embedded star formation in the extended narrow line region of Centaurus A: Extreme mixing observed by MUSE. Astronomy and Astrophysics 590, 37-44.

Scannapieco, E. 2017. The Production of Cold Gas Within Galaxy Outflows. The Astrophysical Journal 837, 28-45.

Schaye, J., Crain, R.A.; Bower, R.G.; Furlong, M.; Schaller, M.; Theuns, T.; et al. 2015. The EAGLE project: simulating the evolution and assembly of galaxies and their environments. Monthly Notices of the Royal Astronomical Society 446, 521-554.

Sijacki, D., Vogelsberger, M., Genel, S., Springel, V., Torrey, P., Snyder, G.F., Nelson, D., Hernquist, L. 2015. The Illustris simulation: the evolving population of black holes across cosmic time. Monthly Notices of the Royal Astronomical Society 452, 575-596.

Silk, J., Rees, M.J. 1998. Quasars and galaxy formation.\ Astronomy and Astrophysics 331, L1-L4.

Stern, J., Faucher-Giguère, C.-A., Zakamska, N.L., Hennawi, J.F. 2016. Constraining the Dynamical Importance of Hot Gas and Radiation Pressure in Quasar Outflows Using Emission Line Ratios. The Astrophysical Journal 819, 130-150.

Sturm, E., González-Alfonso, E.; Veilleux, S.; Fischer, J.; Graciá-Carpio, J.; Hailey-Dunsheath, S.; et al. 2011. Massive Molecular Outflows and Negative Feedback in ULIRGs Observed by Herschel-PACS. The Astrophysical Journal 733, L16-L21.

Sun, A.-L., Greene, J.E., Zakamska, N.L. 2017. Sizes and Kinematics of Extended Narrow-line Regions in Luminous Obscured AGN Selected by Broadband Images. The Astrophysical Journal 835, 222-248.

Tadhunter, C., Morganti, R., Rose, M., Oonk, J.B.R., Oosterloo, T. 2014. Jet acceleration of the fast molecular outflows in the Seyfert galaxy IC 5063. Nature 511, 440-443.

Tadhunter, C. 2008. The importance of sub-relativistic outflows in AGN host galaxies . Memorie della Società Astronomica Italiana 79, 1205.

Tombesi, F., Meléndez, M., Veilleux, S., Reeves, J.~N., González-Alfonso, E., Reynolds, C.S. 2015. Wind from the black-hole accretion disk driving a molecular outflow in an active galaxy. Nature 519, 436-438.

Tombesi, F., Cappi, M., Reeves, J.~N., Braito, V. 2012.\Evidence for ultrafast outflows in radio-quiet AGNs - III. Location and energetics. Monthly Notices of the Royal Astronomical Society 422, L1-L5.

Tombesi, F., Cappi, M., Reeves, J.N., Palumbo, G.G.C., Braito, V., Dadina, M. 2011. Evidence for Ultra-fast Outflows in Radio-quiet Active Galactic Nuclei. II. Detailed Photoionization Modeling of Fe K-shell Absorption Lines. The Astrophysical Journal 742, 44.

Tremblay, G.R., Oonk, J. B. R.; Combes, F.; Salomé, P.; O'Dea, C.P.; Baum, S. A. et al. 2016. Cold, clumpy accretion onto an active supermassive black hole. Nature 534, 218-221.

Veilleux, S., Bolatto, A., Tombesi, F., Meléndez, M., Sturm, E., González-Alfonso, E., Fischer, J., Rupke, D.S.N. 2017. Quasar Feedback in the Ultraluminous Infrared Galaxy F11119+3257: Connecting the Accretion Disk Wind with the Large-scale Molecular Outflow. The Astrophysical Journal 843, 18-29.

Veilleux, S., Meléndez, M.; Sturm, E.; Gracia-Carpio, J.; Fischer, J.; González-Alfonso, E. et al. 2013. Fast Molecular Outflows in Luminous Galaxy Mergers: Evidence for Quasar Feedback from Herschel. The Astrophysical Journal 776, 27-48.

Veilleux, S., Cecil, G., Bland-Hawthorn, J. 2005. Galactic Winds. Annual Review of Astronomy and Astrophysics 43, 769-826.

Zakamska, N.L., Greene, J.E. 2014. Quasar feedback and the origin of radio emission in radio-quiet quasars. Monthly Notices of the Royal Astronomical Society 442, 784-804.

Zubovas, K., King, A.R. 2014. Galaxy-wide outflows: cold gas and star formation at high speeds. Monthly Notices of the Royal Astronomical Society 439, 400-406.

Zubovas, K., King, A. 2012. Clearing Out a Galaxy. The Astrophysical Journal 745, L34.

Wagner, A.Y., Umemura, M., Bicknell, G.~V. 2013. Ultrafast Outflows: Galaxy-scale Active Galactic Nucleus Feedback. The Astrophysical Journal 763, L18.

Wagner, A.Y., Bicknell, G.~V., Umemura, M. 2012. Driving Outflows with Relativistic Jets and the Dependence of Active Galactic Nucleus Feedback Efficiency on Interstellar Medium Inhomogeneity. The Astrophysical Journal 757, 136.

Whittle, M. 1985. The narrow line region of active galaxies. I - Nuclear forbidden line profiles. II - Relations between forbidden line profile shape and other properties. Monthly Notices of the Royal Astronomical Society 213, 1-31.